\documentclass{aa}
\usepackage{graphicx}
\usepackage{txfonts}

\def\lya{Ly$\alpha$~}
\def\lyb{Ly$\beta$~}

\def\hb{H$\beta$~}
\def\xha{H$\alpha$}

\def\hahb{\xha/\hb}

\def\sm{$\cal M_\odot$~}
\def\sma{$\cal M_\odot$}
\def\ma{$\cal M$}

\def\kms{km s$^{-1}$}

\begin{document}

\title{First detection of Lyman continuum photon escape from a local starburst 
galaxy}
\subtitle{Observations of the luminous blue compact galaxy Haro 11 with the Far 
Ultraviolet Spectroscopic Explorer (FUSE). I. 
\thanks{Based on observations made with the NASA-CNES-CSA Far Ultraviolet 
Spectroscopic Explorer. FUSE is operated for NASA by the
Johns Hopkins University under NASA contract NAS-32985}}

\author{ Nils Bergvall\inst{1}
\and 
E. Zackrisson\inst{2}
\and
B-G Andersson\inst{3}
\and
D. Arnberg\inst{4}
\and 
J. Masegosa\inst{5}
\and
G\"oran \"Ostlin\inst{6}
}
\offprints{N. Bergvall}

\institute{Dept. of Astronomy and Space Physics, Box 515, S-75120 Uppsala, 
Sweden 
\\
\email{nils.bergvall@astro.uu.se}
\and
Dept. of Astronomy and Space Physics, Box 515, S-75120 Uppsala, 
Sweden 
\\
\email{erik.zackrisson@astro.uu.se}
\and
Department of Physics and Astronomy, Johns Hopkins University, 3400 North 
Charles Street, Baltimore, MD 21218, USA
\\
\email{bg@pha.jhu.edu}
\and
Dept. of Astronomy and Space Physics, Box 515, S-75120 Uppsala, 
Sweden 
\and
Instituto de Astrofisica de Andalucia, Granada, Spain
\\
\email{pepa@iaa.es}
\and
Stockholm Observatory, SCFAB, SE-106 91 Stockholm, Sweden 
\\
\email{ostlin@astro.su.se}}

\date{Received 7 July 2005 / Accepted 2 November 2005}

\abstract{The dominating reionization source in the young universe has not yet 
been identified. Possible candidates include metal poor dwarf galaxies with 
starburst properties.}{We selected an extreme starburst dwarf, the Blue Compact 
Galaxy Haro 11, with the aim of determining the Lyman continuum 
escape fraction from UV spectroscopy.}
{Spectra of Haro 11 were obtained with the Far Ultraviolet 
Spectroscopic Explorer (FUSE). A weak signal shortwards of the Lyman break is 
identified as Lyman continuum (LyC) emission escaping from the ongoing 
starburst. From profile fitting to weak metal lines we derive column densities 
of the low ionization species. Adopting a metallicity typical of the H~II 
regions of Haro 11, these data correspond to a hydrogen column density of 
$\sim$10$^{19}$cm$^{-2}$. This relatively high value indicates that most of the 
LyC photons escape through transparent holes in the interstellar medium. We then 
use spectral evolutionary models to constrain the escape fraction of the 
produced LyC photons.}
{Assuming a normal Salpeter initial mass function we obtain a Lyman continuum escape fraction 
of f$_{esc}\sim$ 4--10\%. We argue that in a 
hierarchical galaxy formation scenario, the upper limit we derive for the escape 
rate allows for a substantial contribution to cosmic reionization by starburst 
dwarf galaxies at high redshifts.}
{}


\maketitle

\section{Introduction}

According to WMAP data, the epoch of reionization started at a redshift of 
$z\approx 20$ (Kogut et al. \cite{kogut}). Observations of the high opacity in 
the Lyman continuum (hereafter LyC) emission of galaxies (Becker et al., 
\cite{becker}, Fan et al. \cite{fan2}) and the results from the recent studies 
of the intergalactic medium (IGM) around SDSS quasars (e.g. Mesinger \& Haiman 
\cite{mesinger}) indicate that the final stage of the epoch of reionization may 
have occurred at redshifts $z\sim$6--7. There is still no consensus about what 
could have been the major ionization sources of the IGM during the reionization 
epoch -- stars, normal galaxies, AGNs or some exotic mechanism (Yan \& Windhorst 
\cite{yan}, Meiksin \& White \cite{meiksin}). The rapid decline of the space 
density of quasars at redshifts $z>$3 (Madau et al. \cite{madau1}, Fan et al. 
\cite{fan1}) makes it likely that their contribution to the LyC production rate 
required for the reionization is insignificant, although it has been proposed 
that miniquasars fed by intermediate-mass black holes may have been important 
(Madau et al. \cite{madau2}). 

At lower redshifts, observations of a sample of Lyman-break galaxies at $z \sim$ 
3.4 (Steidel et al. \cite{steidel}) seem to show that galaxies could account for 
a significant fraction of the photon flux required for reionization. The 
observations indicate a contribution of LyC photons a factor of $\sim$5 
higher than for quasars at the same epoch. These results have however not been 
confirmed by other similar studies. On the contrary, it has been argued 
(Fern\'andez-Soto et al. \cite{fernandez-soto}) that a maximum of 4\% of the LyC 
flux could escape from luminous galaxies at redshifts 1.9$<z<$3.5. This is also 
close to the upper limit of the escape fraction of low redshift galaxies 
(Leitherer et al. \cite{leitherer1}; Hurwitz et al. \cite{hurwitz}; Deharveng et 
al. \cite{Deharveng et al.}; Heckman et al. \cite{heckman}). The lack of a 
strong trend with redshift indicates that galaxies at higher redshifts also have 
low escape fractions, insufficient to keep the intergalactic gas ionized. 
Alternative ionization sources that have been discussed are massive pregalactic 
Pop III stars and young globular clusters. Most probably the reionization is 
ruled by different sources at different redshifts. It is therefore important to 
establish observational constraints on the LyC fluxes from the various 
candidates at different epochs. 

Our information about the galaxy luminosity/mass function at the epoch of 
reionization is very meager, and it is not at all clear what mass scale would 
dominate the LyC photon production rate. In a galaxy formation scenario where 
mergers are of major importance one would expect starburst dwarf galaxies to be 
frequent among the first generation of galaxies. Indeed the number density of so 
called Luminous Compact Blue Galaxies (LCBGs) appears to increase with redshift 
(Lilly et al. \cite{lilly}; Mall\'en-Ornelas et al. \cite{mallen-ornelas}). It 
is therefore important to investigate if this type of galaxy could have been an 
important ionization source at high redshift. Dwarf galaxies are however too 
faint to be observed at high 
redshifts. We are therefore dependent on theoretical models and observations 
of nearby galaxies that can be used as templates of the first generation of 
dwarf galaxies. Recent models do not give much hope for Lyman escape from 
galaxies of masses larger than 10$^7$ solar masses (Kitayama et al. 
\cite{kitayama}) but these predictions have to be tested observationally. 

We have searched for local dwarfs with properties similar to young dwarf 
galaxies, that could be used as test objects. A few such studies have already 
been carried out but only provided upper limits to the escape fraction 
(e.g., Giallongo et al. \cite{giallongo}; Devriendt et al. \cite{devriendt}; 
Shull et al. \cite{shull}). However, predicted equivalent widths of the Balmer 
emission lines in starbursts are in some cases larger than observed (e.g. 
Bergvall \cite{bergvall1}, Bresolin et al., \cite{bresolin}, Mas-Hesse \& Kunth 
\cite{mashesse}, Moy et al. 
\cite{moy}, Stasinska et al. \cite{stasinska}). Several possible explanations to 
this were discussed by Mas-Hesse \& Kunth (\cite{mashesse}). One possibility is 
that the most massive stars remain hidden in their parental clouds until most of 
the fuel has been consumed. This would mimic a truncation of the IMF at high 
masses. Mas-Hesse and Kunth also discussed the possibility that a large fraction 
of the ionizing photons are absorbed by dust in the H II regions. To obtain 
agreement with the observations, a large amount of the photons, $\sim$ 60\% 
would have to be absorbed. This possibility has recently been investigated by 
Hirashita et al. (\cite{hirashita}) where the escape fraction of the LyC photons 
from the H II region itself is estimated from model comparisons. Based on input 
data from a sample of star forming galaxies they find that, in the mean about 
40\% of the photons are trapped, which is less than what would be needed 
according to Mas-Hesse and Kunth. But the scatter is high so in some cases this 
certainly would be a possibility. Hirashita et al. seem to find that the escape 
fraction is higher is galaxies with properties similar to Haro 11 (e.g. the BCG 
ESO 338-IG04). Therefore an alternative, or complementary, explanation for the 
low flux of the recombination lines, may be that a significant fraction of the 
LyC photons is in fact escaping from the galaxy. This is the possibility that we 
study in this paper. 

H~II regions in nearby galaxies appear to leak 40--50\% of the LyC photons 
produced by the hot stars into the diffuse interstellar medium (Ferguson et al. 
\cite{ferguson}; Oey et al. \cite{oey}). Although it cannot be completely 
excluded that the gas may be ionized by single O stars located in situ, a 
comparison of the observed emission line strengths with theoretical models 
(Iglesias-P\'aramo \& Mu\~noz-Tu\~n\'on \cite{iglesias-paramo}; Wood \& Mathis 
\cite{wood}) give support to the leakage interpretation. An even more extreme 
case of a density bounded H~II region around a massive star cluster was reported 
by Leitherer et al. (\cite{leitherer2}). They estimate that about 75\% of the 
LyC radiation is leaking into the diffuse ambient medium. Depending on the 
distribution of the gas, the location where these photons are absorbed may be 
very different from case to case. In some situations the diffuse ionized gas may 
be much more extended 
than the H~II region producing the photons and will be ignored by the observer 
who consequently will underestimate the star formation rate.

The other alternative for the discrepancy between theory and observations of the 
Balmer emission line strengths is that the dust is unevenly distributed across 
the starburst region so that the younger regions are more reddened than the 
older ones (Calzetti et al. \cite{calzetti}). This could reduce the equivalent 
width of H$\alpha$ with up to a factor of $\approx 2$.  The only way to really 
distinguish between a global leakage into to intergalactic medium and the other 
two alternatives is to search for leaking photons shortward of the Lyman limit.

In previous studies of starburst dwarfs (Bergvall \cite{bergvall1}, Bergvall \& 
\"Ostlin \cite{bergvall4}, \"Ostlin et al. \cite{ostlin1}, \cite{ostlin2}) we 
have focused on a few luminous (--19$> $M$_B>$--21) metal poor blue compact 
galaxies (BCGs). They have properties similar to the LBGs at intermediate 
redshifts (e.g. Guzman et al. \cite{guzman}). Here we report on such a study 
using FUSE, the Far Ultraviolet Spectroscopic Explorer.

\subsection{The target galaxy}

The target galaxy of our choice is Haro 11 (=ESO 350-IG38) at 
$\alpha_{2000}$=00$^h$36$^m$52.5$^s$, 
$\delta_{2000}$=-33$^{\circ}$49$\arcmin$49$\arcsec$. From the emission-line 
spectrum we derive a heliocentric radial velocity of v$_0$=6180 km s$^{-1}$, 
corresponding to a distance of 86 Mpc, assuming a Hubble constant of 72 km 
s$^{-1}$ Mpc$^{-1}$. We have previously studied 
this galaxy over a 
wide frequency range, from UV to radio (Bergvall \& Olofsson \cite{bergvall2}, 
Bergvall et al. \cite{bergvall3}, \cite{bergvall4}, \"Ostlin et al., 
\cite{ostlin1}, \cite{ostlin2}). It has unique properties that makes it one of 
the most relevant local objects for our purpose. The chemical abundances are 
low, the oxygen abundance being $\sim$20\% solar (relative to the revised solar 
oxygen abundance by Asplund et al. \cite{asplund}). 

An estimate of the typical hydrogen mass of a BCG can be obtained from the 
correlation between gas mass and major diameter (Gordon \& Gottesman 
\cite{gordon}). Haro 11 has a major diameter of 11 kpc at $\mu_B$=25 
mag arcsec$^{-2}$ indicating that the H~I mass should be approximately 
2$\times$10$^9$ \sma. Despite several efforts at VLA, Nancay and Parkes (all 
unpublished) we have not been able to detect H~I in this galaxy. We estimate the 
upper limit to 
\ma (H~I)$\sim$ 10$^8$ \sma. The extraordinary strong [C~II]$\lambda$158$\mu$ 
line, observed with ISO (Bergvall et al. \cite{bergvall3}), shows that a large 
part of the H~I gas is located in photodissociation regions (PDRs). We have 
estimated the contribution from H$_2$ and H~II gas to between 10$^8$ and 10$^9$ 
\sma~ each. This would leave a remarkably small fraction of the total gas mass 
in atomic form. A possible explanation is that the bulk of the gas in the halo 
that normally is in neutral state has become ionized by the starburst. Our 
H$\alpha$ observations show that the main body is indeed surrounded by an 
extended halo of ionized gas. The conditions for LyC photon escape thus appear 
to be unusually favourable. It is interesting to note that the nearby BCG Pox 
186, sometimes called 'ultracompact', is also H I quiet (Begum \& Chengalur 
\cite{begum}).

HST GHRS UV spectroscopic observations of Haro 11 were carried out by Kunth et 
al. (\cite{kunth}) who found a quite complex absorption system with velocity 
components of Ly$\alpha$ in absorption on both sides of the emission line. Thus 
it appears that there are both outflows and infall of gas in the centre 
direction. From the absorption lines Kunth et al. estimate column densities and 
find values as high as log n$_H \sim$ 20 cm$^{-2}$. The aperture used is small 
however (2$\arcsec$ circular diameter) and covers only the very central part of 
the galaxy. Our optical spectroscopy (\"Ostlin et al. \cite{ostlin05}) reveals 
velocity gradients of both gas and stars as large as 200 km s$^{-1}$ across two 
of the bright condensations in the very centre, indicating an extremely high 
mass density in this region.

\section{Observations and reductions}

\subsection{IUE observations}

\begin{figure}
\resizebox{\hsize}{!}{\includegraphics*{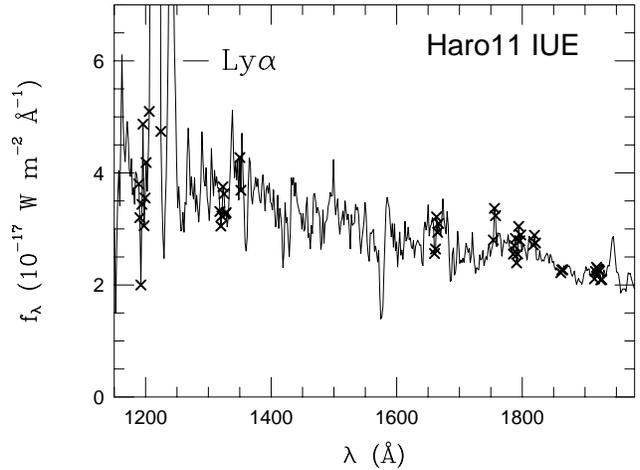}}
\caption[]{Spectrum of Haro 11 obtained with the IUE spacecraft in the low 
resolution SWP mode. The aperture used was 10$\arcsec \times$20$\arcsec$. Bad 
pixels have been flagged with crosses.}
\label{3788fig1}
\end{figure}

In 1984 we observed the galaxy in low dispersion mode with the SWP UV 
spectrograph of the IUE spacecraft. The aperture used was 
10$\arcsec$x20$\arcsec$. A low resolution spectrum is shown in Fig. 
\ref{3788fig1}. 
Below we will use the slope of the spectrum as an input to the derivation of the 
dust extinction. 

We will also use the  IUE spectrum to determine the C/O abundance ratio, based 
on the C III$\rbrack\lambda 1908$, and O III$\rbrack\lambda 1663$ lines. The 
strongest line, also representing the dominating ionization stage is C 
III$\rbrack\lambda 1908$ while the O III$\rbrack$ line is nosier. Thus the error 
in the determination will be significant but it still will be useful as a check  
of the C/O ratio determined from the FUSE absorption lines (see sect. 3.1). The 
atomic collision strengths are determined from the electron temperature derived 
by Bergvall \& \"Ostlin (\cite{bergvall4}) $T\approx 13400$ K. From the relative 
line intensities we then obtain N(C$^+$)/N(O$^{++})\sim 0.3 \pm 0.1$. The error 
is dominated by the photon statistics and has been estimated from the noise in 
the region around the line. Assuming ionization correction factors  
ICF(C/C$^+$)$\sim$ 1.4 (Stasi\'nska \cite{stasinska2}) and ICF(O/O$^{++}$)$\sim$ 
1.5 (Bergvall \& \"Ostlin \cite{bergvall4}), we then finally obtain 
N(C)/N(O)=$0.46^{+0.15}_{-0.10}$.

\subsection{FUSE observations}

The data were obtained on Oct 12, 2001, using the FUSE low resolution mode LWRS 
with an aperture of 30$\arcsec$x30$\arcsec$. The total integration time on 
target was 16ks. 12ks of the observations were carried out during orbital night. 
The data were reduced using CalFUSE v3.0, with default processing parameter for 
all spectral segments. As the background level in FUSE data is quite low and 
hence cannot be reliably measured in detail for short observations, the standard 
FUSE background correction is based on a set of template files which are scaled 
to averages over the whole detector in any given observation. These template 
files consist of coadded data from many, long observations of blank sky acquired 
throughout the mission. The FUSE detectors 
consist of microchannel plates with electronic pixels whose 
exact locations depend on the electrical field structure and read-out 
electronics of the detector (Sahnow et al., 2000). Artificial count 
enhancements occur at the edges of the detector due to non-uniformities in 
the field and have a very steep dependence on the distance from the edge.  The 
exact mapping of the detector to location in the electronic detector-image 
depends on several factors including the total brightness of the detector 
illumination.  Because of the steepness of this edge effect, minor mismatches in 
the target and background images can cause significant errors in the background 
subtraction (Sahnow, 2005, private communication). As the 
SiC spectrum falls very close to the lower edge of detector 1B, any relative 
change in the strength of this edge effect compared to the rest of the 
background will cause problems in the background correction, particularly for 
weak sources such as Haro 11. To compensate for this effect, we reset the areas 
and relative weighting used to determine the scaling of the background template 
in the reduction of the SiC1B data.  Without any further modifications the 
derived SiC1B spectrum then agrees well with the SiC2A spectrum. The regions of 
the lowest signal in both the SiC1B and the LiF1A segments were however 
significantly below the zero level, indicating that the proper background on the 
whole was fainter than the subtracted template background. After extraction, 
calibration and combination of the individual segments, a constant was therefore 
added to the background to compensate for this. As we are more strongly 
constrained by low signal-to-noise than spectral resolution, we are not 
significantly affected by the slight difference in dispersion solution between 
segments, and the consequent lowering of resolution caused by this coaddition.

\section{Results}

\begin{figure}[h]
\resizebox{\hsize}{!}{\includegraphics*{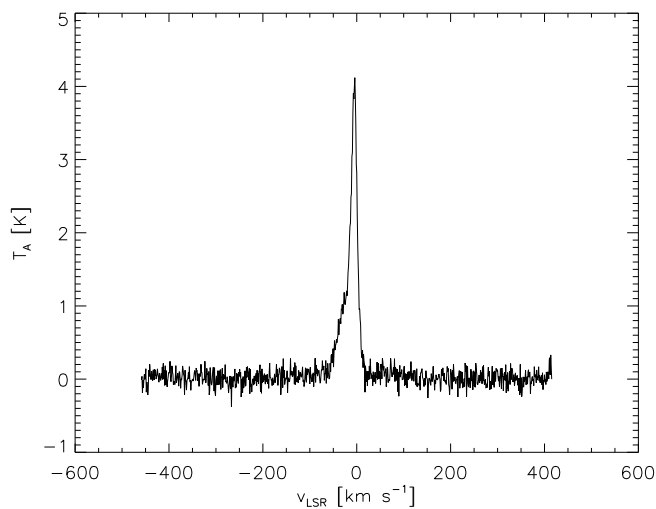}}
\caption[]{The 21-m line in the direction of Haro 11 obtained from the 
Leiden/Dwingeloo survey (Burton \& Hartmann \cite{burton}). V$_{LSR}$ is the 
radial velocity relative to the local standard of rest and T$_A$ is the antenna 
temperature.}
\label{3788fig2}
\end{figure}

\begin{figure*}
\resizebox{\hsize}{15cm}{\includegraphics*{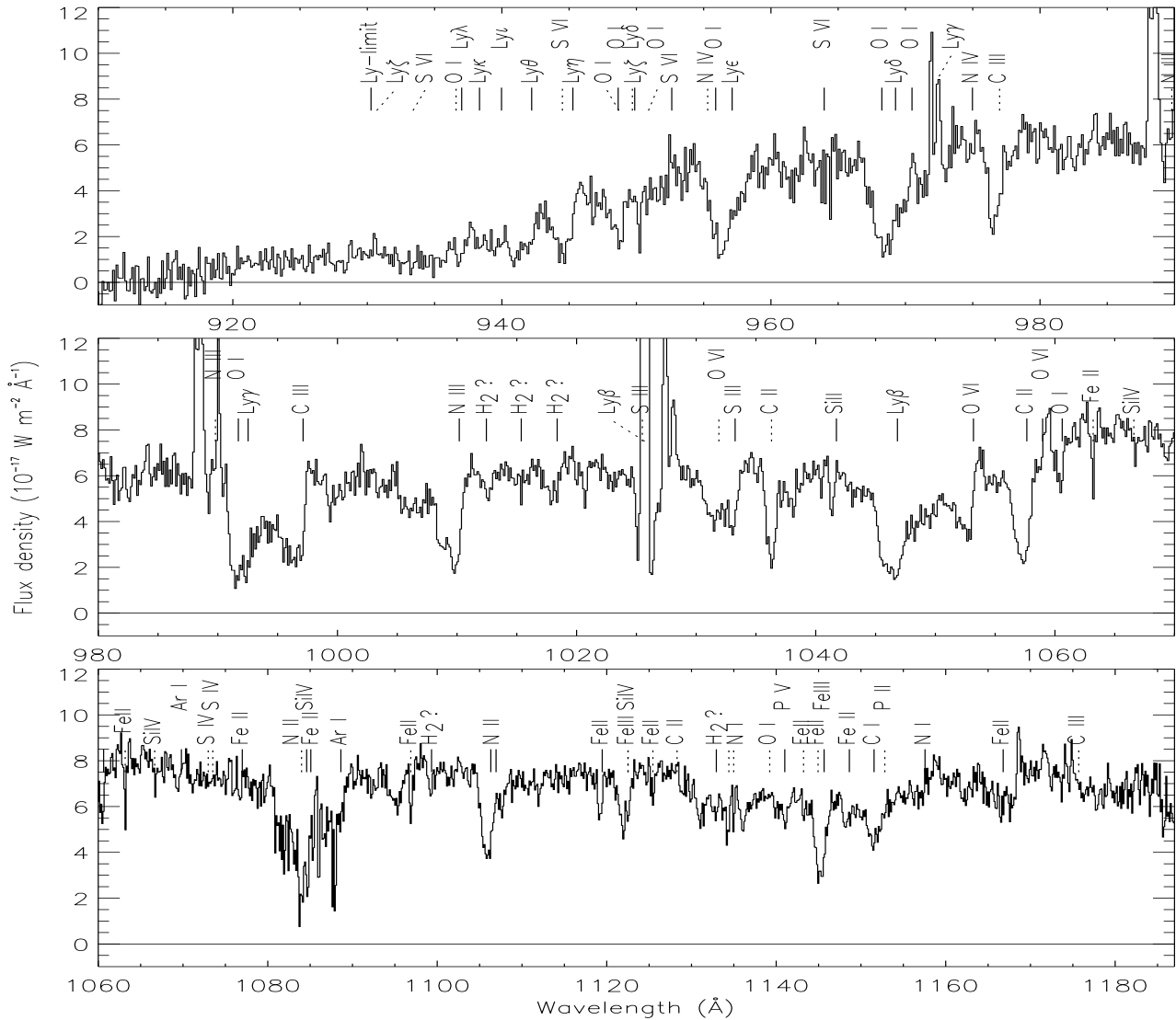}}
\caption[]{FUSE spectrum of Haro11. The major absorption features have been 
indicated with solid lines (Haro 11 optical rest frame) and hatched lines (local 
frame). The spectrum also shows strong geocoronal Lyman emission lines and 
scattered solar HeI at $\lambda$1169 \AA. The data have been rebinned with a 
boxcar of size 0.26 \AA.}
\label{3788fig3}
\end{figure*}

Fig. \ref{3788fig3} shows the full spectrum with a few of the strongest 
absorption 
and emission lines indicated. Strong geocoronal Lyman emission 
lines are seen on top of the absorption spectrum. The narrow spike at 1168.7 \AA 
~is a scattered HeI solar emission line. From the 
absorption lines we can define two velocity systems, one originating in the 
Milky Way and the other in Haro 11. Fig. \ref{3788fig2} shows the H~I 21-cm 
line spectrum in the 
direction of Haro 11 (l=328$^\circ$, b=-83$^\circ$), obtained from the 
Leiden/Dwingeloo survey (Burton \& Hartmann \cite{burton}). 
Our analysis shows that the spectrum in the velocity range -450 \kms -- 400 \kms 
can be approximated with two components with small negative velocities relative 
to the LSR  as shown in Table \ref{dwingelootab}.

\begin{table}[h]
\caption[]{Components of the local H~I spectrum in the direction of Haro 11. The 
columns show the central velocity relative to the local standard of rest, peak 
antenna temperature, full width at half maximum and the derived column density. 
}
\begin{flushleft}
\begin{tabular}{llll} 
V(centre) & T(peak) & FWHM & N \cr
\kms & K & \kms & cm$^{-2}$ \cr
\hline
-5.1 & 3.1 & 13.4 & 8.1$\times$10$^{19}$ \cr 
-19.8 & 1.5 & 42.6 & 9.5$\times$10$^{19}$ \cr 
\hline
\end{tabular}\\
\label{dwingelootab}
\end{flushleft}
\end{table}

The spectral features identified to belong to the Milky Way ISM are indicated 
with hatched 
vertical lines in Fig. \ref{3788fig3}. These absorption lines have a 
linewidth 
of FWHM $\sim$ 30 km s$^{-1}$ in agreement with the FWHM given in Table 
\ref{dwingelootab}. The narrowness of the lines is important since it ensures 
that we can find the proper level of the Lyman continuum of Haro 11 between the 
positions of the galactic Lyman absorption lines.
The spectrum of Haro 11 is rich in spectral features and dominated by absorption 
lines from O and B stars. Possible detections of molecular hydrogen lines in the 
Lyman series have also been indicated. The heliocentric systemic velocity of 
Haro 11, as defined from the optical H~II emission (unpublished) and the stellar 
CaII triplet absorption lines, weighted from the UV mapping of the galaxy 
(\"Ostlin et al. \cite{ostlin05}), is v=6180 km s$^{-1}$. This velocity agrees 
with the position of the P V $\lambda$1118 \AA~($\lambda_{obs}$=1141 \AA) line. 
The P V line originates from hot dwarf stars on the main sequence. If these 
stars are metal-rich and massive they tend to show P Cygni profiles while at 
lower metallicities, as in our case, the feature is mainly photospheric and 
narrow and can be used as a systemic velocity fixpoint. Thus we have two 
independent derivations of the systemic velocity of the stellar population seen 
in the far-UV spectral region. Strong deviations from this value would indicate 
in- or outflows of gas. 

The strongest absorption lines, like those of the Lyman series, are dark but not 
saturated as is normally the case for starburst dwarfs (e.g. Heckman et al. 
\cite{heckman}). This in itself is an indication that the optical depth in the 
LyC is lower than normal for this type of galaxies. Many of the absorption lines 
originating in Haro 11 are broad and show P Cygni profiles, e.g. the O VI 
$\lambda \lambda$1032,1038 (redshifted to $\lambda$1053,1058 \AA) lines. Most of 
these lines are probably formed in the photospheres or in the winds of O and B 
stars but there may also be a contribution from more global outflows. 
After a rough correction for the stellar contribution (see next section), some 
of the strongest absorption features show asymmetries, indicating high-velocity 
outflows. This aspect will be briefly discussed below and more in detail in a 
forthcoming paper, focused on the absorption line spectrum.

   \begin{figure}[h]
	\resizebox{\hsize}{!}{\includegraphics*{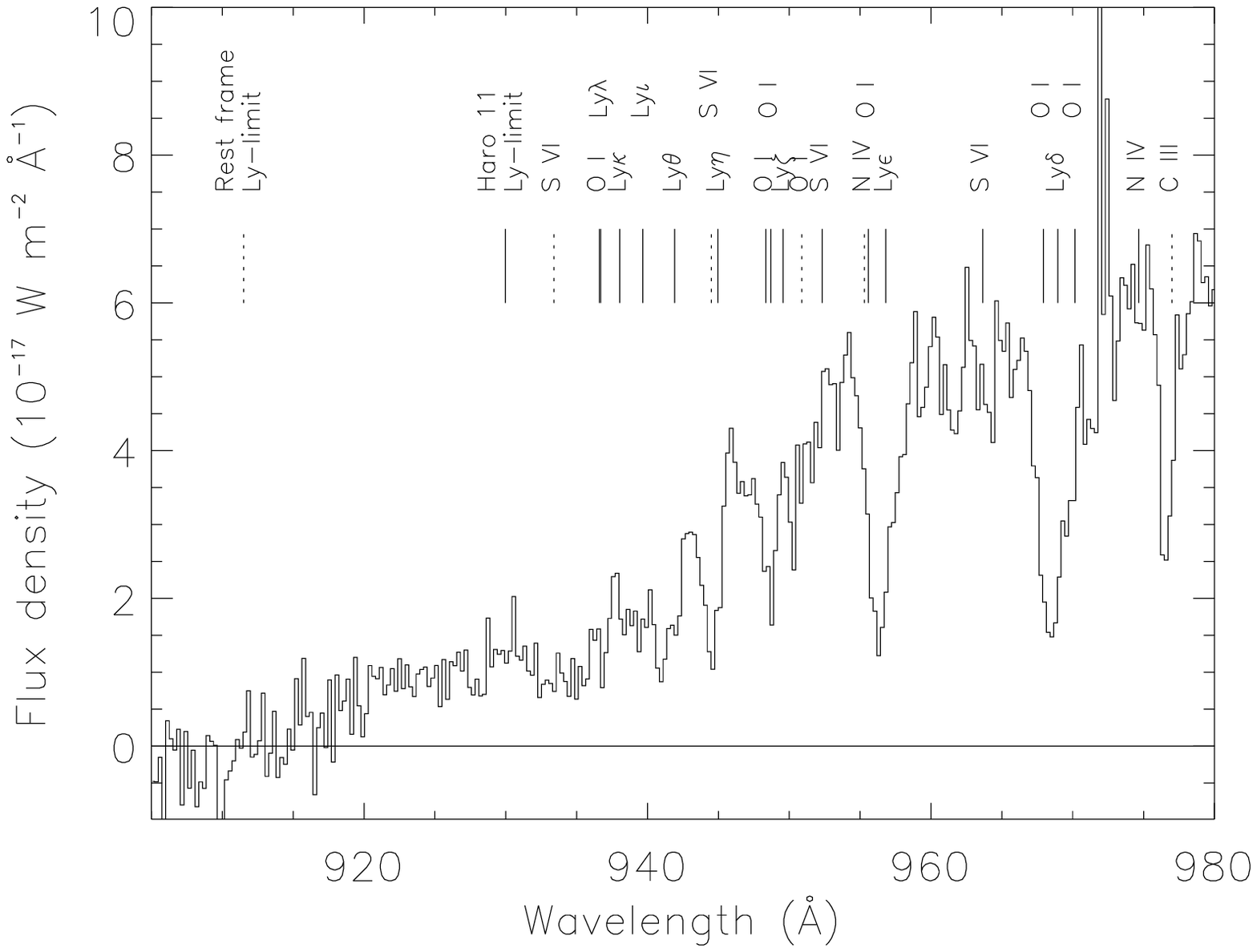}}
      \caption[]{Part of spectrum showing the Lyman continuum signal. The Lyman 
limit in the local frame (hatched lines) and in the frame of Haro 11 (solid 
lines) have been indicated.}
\label{3788fig4}
   \end{figure}

   \begin{figure}[h]
	\resizebox{\hsize}{!}{\includegraphics*{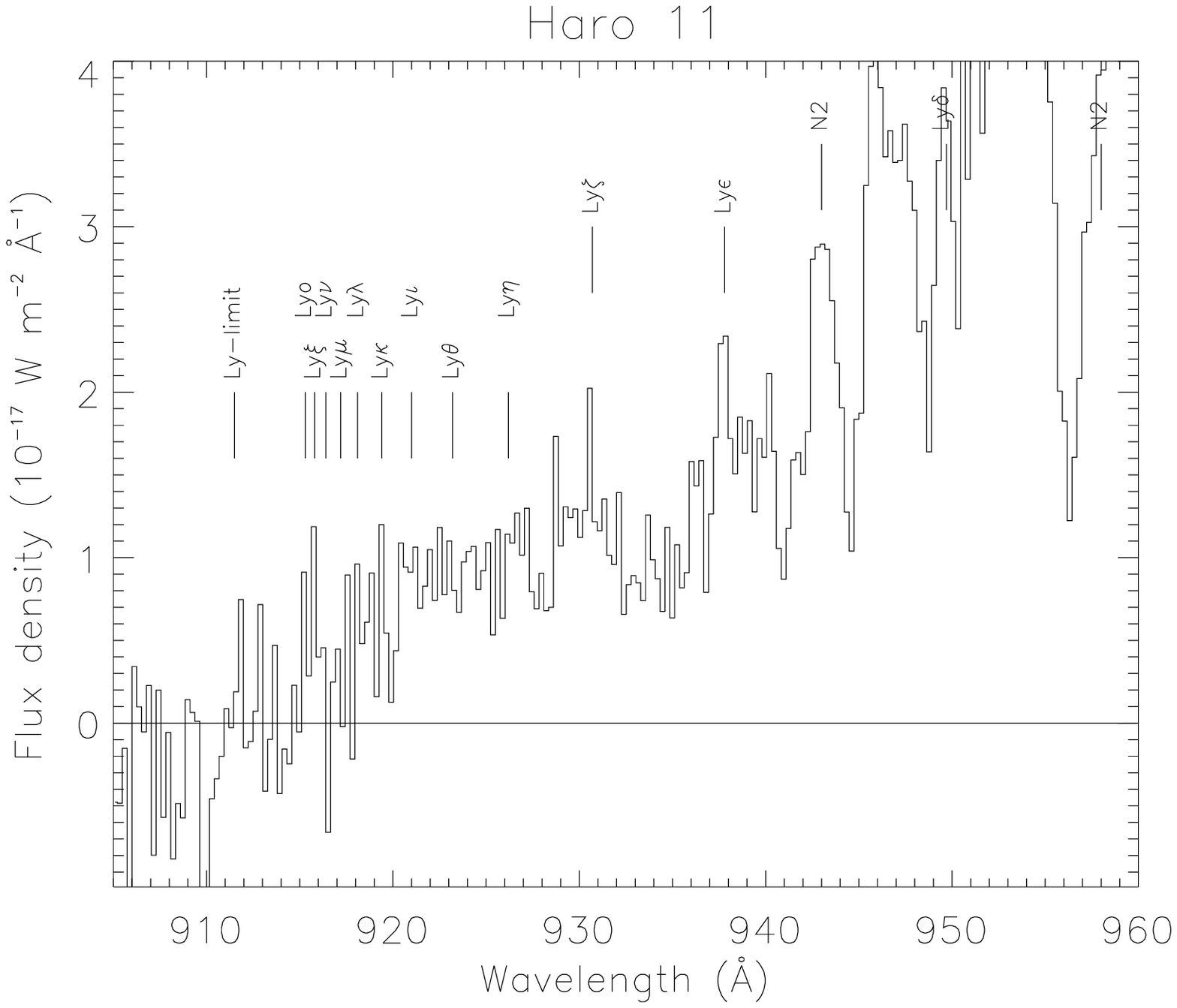}}
      \caption[]{Airglow lines and the rest frame Lyman series.}
\label{3788fig5}
   \end{figure}

In Fig. \ref{3788fig4} we display the short wavelength part of the spectrum. 
This 
spectral region is covered by two detector segments, 1B and 2A, producing two 
independent spectra SiC1B and SiC2A. Wavelengths below 917 \AA~ are only covered 
by segment 1B.  This is also the detector segment with the smallest effective 
area of the two. Thus, in this region the signal is noisier than at longer 
wavelengths. What is particularly interesting from Fig. \ref{3788fig4} is that 
below the Lyman limit 
of Haro 11 there clearly seems to be {\it a weak but significant signal in the 
Lyman continuum}. Shortwards of the Lyman limit of the local rest frame the 
signal diminishes gradually, as expected from the increasing density of the 
Lyman absorption lines of the local ISM. The emission is seen in the 
2-dimensional display of the spectral region as a faint narrow noisy structure 
in both SiC channels. We therefore feel confident that the LyC excess is not due 
to calibration problems.  The level of the signal however is not so well 
determined, mainly because of uncertainties in the calibration, a low signal and 
possibly influences of auroral emission lines and emission and absorption lines 
from the Milky Way which we will now discuss.

   \begin{figure*}
	\resizebox{\hsize}{!}{\includegraphics*{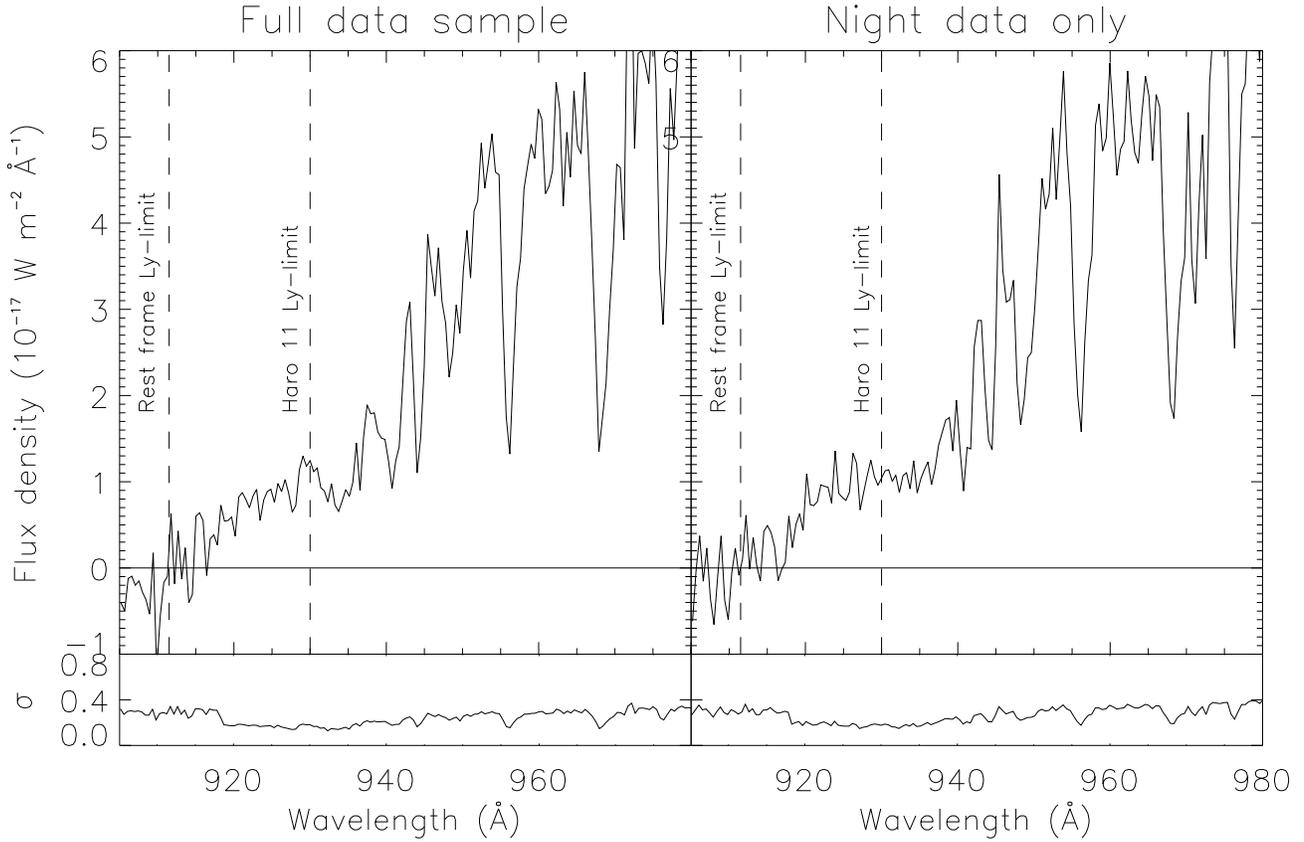}}
      \caption[]{Comparison between night and day data. The lower panel shows 
the 1 $\sigma$ noise level, corrected to the binsize used in the plot.}
\label{3788fig6}
   \end{figure*}


Haro 11 is at a galactic latitude of -83 deg. and thus interstellar absorption 
is not a serious problem. From Schlegel at al. (\cite{schlegel}) we obtain 
A$_B$=0.049$^m$ corresponding to A$_{Ly-limit}\sim$0.18$^m$ (Mathis 
\cite{mathis}). The Lyman emission lines from the terrestrial 
airglow are indicated in Fig. \ref{3788fig5}. For comparison we show in Fig. 
\ref{3788fig6} both the spectrum based on night-and-day data and the spectrum 
based on night data only. We can clearly identify the geocoronal Lyman emission 
lines. Shortwards of the Lyman lines, at much 
lower intensities, C~I lines may weakly influence the counts (Feldman et al. 
\cite{feldman}). In the following discussion we 
will use the 900 \AA~ rest frame flux density of Haro 11 as a reference of the 
LyC 
level. This choice allows comparisons with previous studies of other galaxies 
(e.g. Leitherer et al. \cite{leitherer1}, Deharveng et al. \cite{Deharveng et 
al.}, Inoue et al. \cite{inoue}). 
In the local frame this part of the spectrum has been redshifted to 
$\lambda$=920 \AA. To beat down the noise in the spectrum we determine the LyC 
level over the line-free regions in the interval 920-925 \AA, since the LyC 
signal is not expected to vary significantly over such a short interval. After 
correcting for galactic extinction, we obtain 
f$_{900}$=1.1$\pm$0.1$\times$10$^{-17}$Wm$^{-2}$\AA$^{-1}$.
The uncertainty due 
to the complexity in the determination of the background is probably 
significantly larger than the formal error. In the worst case it could amount to 
of the order of $\sim$ 50\% of the signal.

\subsection{Spatial distribution of the absorbers.}

\begin{figure}
\resizebox{\hsize}{!}{\includegraphics*{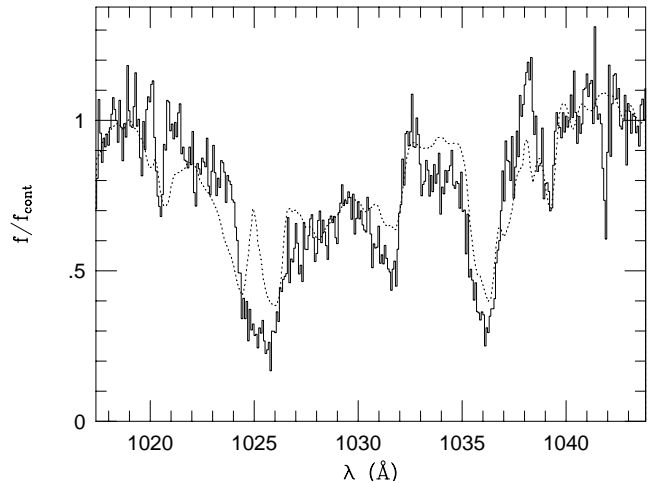}}
\caption[]{The FUSE spectrum of Haro 11 (solid line) along with the best fitting 
stellar spectrum 
(dotted line) from Robert et al. (\cite{robert}). The wavelength scale has been 
corrected to the 
rest frame of Haro11, assuming a systemic heliocentric velocity of 6180 km 
s$^{-1}$. The model 
spectrum is based on a Salpeter IMF, a stellar mass range of 1--120 \sma, a 
metallicity of 18\% 
solar and a continuous SFR. The age is 14 Myr.}
\label{3788fig7}
\end{figure}

We mentioned in the introduction that Haro 11 appears to be devoid of neutral 
hydrogen. This could indicate that the LyC photons are leaking out from a 
truncated Str\"omgren sphere, i.e. that the emission is 'density bounded'. The 
hydrogen ionization cross section at the Lyman limit is 
$\approx 8\times 10^{-18}$cm$^{2}$. This means that the LyC is essentially 
black at column 
densities of a few times 10$^{17}$cm$^{-2}$, if the neutral gas is diffusely 
distributed. As discussed in sect. 1.1 we have derived an upper limit of the 
mass of  
neutral hydrogen in Haro 11 
of 10$^8$ \sma. We can use this number to estimate the maximum column density of 
neutral hydrogen under the assumption that the H~I has the same spatial 
distribution as the stars in Haro 11. The stellar scalelength is {\it h$_V 
\sim$} 2.5 kpc (Bergvall \& \"Ostlin \cite{bergvall4}). This gives an
H~I column 
density in the direction of the centre of N(H~I)=5$\times$10$^{19}$ cm$^{-2}$
which can be regarded as an {\it upper limit}, considerably larger than the 
limiting column density for Lyman leakage. In this simple model H~I would become 
transparent at 4-5 scalelengths (i.e. $\sim$ 10 kpc) from the centre, where the 
host galaxy component starts to dominate. But we know that the neutral gas is 
clumped so we will adapt the simplest alternative to the homogeneous model - the 
'picket-fence' model. Here it is assumed that the LyC radiation is leaking out 
through holes in the 
ambient gas. In 
principle we can carry out profile fitting to the Lyman lines in absorption to 
derive the H~I 
column densities. These lines are however not very useful since there is a mixed 
contribution from 
OB star photospheres, ISM absorption and line emission from the H~II regions. We 
can make an 
estimate of the influence of stellar features from models of spectra of young 
stellar populations 
by Robert et al. (\cite{robert}). These models are based on observed stellar 
spectra over a range 
of metallicities. The metal-poor models are valid for ages below 15 Myr. We 
interpolated the 
model spectra to a metallicity in agreement with what we obtained from the 
optical spectra, i.e. 
an oxygen abundance of $\sim$18\% solar (based on the solar oxygen abundance 
derived by Asplund 
et al. \cite{asplund}). We also assumed a Salpeter initial mass function (IMF) 
and a constant star formation rate (SFR). Fig. 
\ref{3788fig7} 
shows the best fit to these models, a burst of the highest age in the library, 
14 Myr. The optical spectra of the central burst show strong W-R features and, 
as we mention above, P Cygni profiles are seen in the FUSE spectrum. These do 
not fit to the model spectrum shown in fig. \ref{3788fig7}. A better fit can be 
obtained if we assume an increasing SFR with the UV emission dominated by 
massive stars of an age of a few Myr. Modifications of the standard IMF can 
further improve the fit. There may also be a significant contribution to the P 
Cygni component from the supernova generated superwind. However, these issues 
are not of main importance here.

The influence of young stars in the spectrum is strong. The presence of O~VI and 
its P Cygni profile indicates the presence of O supergiants (e.g. Gonz\'alez 
Delgado et al. \cite{gonzalez}). Thus the burst must be active or stopped not 
more than 3-4 Myrs ago. For a continuous SFR we do not expect the spectrum to 
change much over the 
expected maximum 
duration of a typical burst, $\sim$ 10$^8$ yr and therefore the comparison in 
Fig. \ref{3788fig7} 
should be qualitatively correct. The model spectrum has only partly been 
corrected for 
interstellar absorption but the influence on the Ly$\beta$ line is marginal so 
the line profile 
should be entirely dominated by the stellar photospheric contribution. This is 
confirmed also from a comparison with predicted \lyb absorption line strengths 
in models of O star spectra (Gonz\'alez Delgado et al. \cite{gonzalez}). Thus it 
is 
obvious that 
these features have 
a strong influence and thus the Lyman lines cannot be used to derive the column 
density without a 
more careful correction for the influence of the stellar features. However, if 
we can derive 
column densities of 
low ionization stages of heavy elements with known abundances relative to 
hydrogen we can obtain 
a stricter constraint on the H~I column density since the atoms and ions in 
these stages to a first approximation should coexist with the neutral hydrogen. 

We will use a few rather well defined and isolated low ionization lines, C~II 
$\lambda$1036, O~I $\lambda$1039, SiII $\lambda$1021, ArI $\lambda$1048, FeII 
$\lambda$1097 and $\lambda$1125. We cannot exclude the fact that there may be a 
weak contribution to some of these lines from the stellar photospheres but the 
bulk of the absorption should originate in the ISM. The ionization 
potentials of these lines are slightly higher than 1 Ryd. It is known from 
studies of the local interstellar medium that elements with ionization 
potentials below that of hydrogen and the ionization potential of the second 
ionization stage above, exist predominantly in the singly ionized stage where 
hydrogen is neutral. It is therefore possible to use also these lines to obtain 
good estimates of N(HI) under certain conditions. The lines of the 
neutral atoms O and Ar can directly be used to derive column densities if the 
atoms are effectively shielded 
from ionizing photons by the neutral hydrogen. In regions where H~I is partially 
ionized, argon may be more fully ionized as a consequence of it high cross 
section to photon absorption. If so, we would underestimate the column density 
based on Ar I. This seems to be confirmed from the results in Table 
\ref{columndensities}. Oxygen behaves differently since the ionization fraction 
is coupled to that of hydrogen via resonant charge-exchange reactions (Jenkins 
et al. \cite{jenkins}). 

The optical depth of a specific species is given by:

\begin{equation}
\tau(\lambda)={{\pi e^2}\over{m_ec^2}}f\lambda^2N(\lambda)
\end{equation} 

where {\it f} is the oscillator strength. From this we can derive the column 
density 
in cm$^{-2}$:

\begin{equation}
N=\int{N(v)dv}={{3.768\cdot 10^{14}}\over{f\lambda}}\int{\tau(v)dv}
\end{equation}

where $\lambda$, the wavelength of the line, is given in \AA~ and v is the 
velocity in 
km~s$^{-1}$. 

\begin{figure}
\resizebox{\hsize}{!}{\includegraphics*{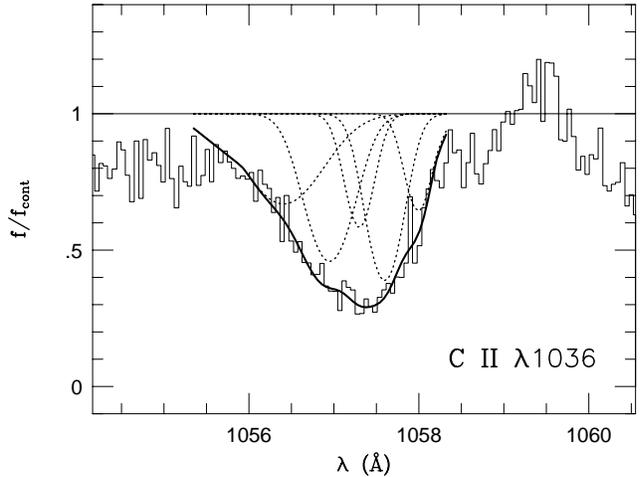}}
\caption[]{The best free fit to the observed C II $\lambda$1036 line based on 5 
velocity components 
with different Doppler widths and component separations. The strongest component 
is centered on the systemic velocity as derived from the optical data.}
\label{3788fig8}
\end{figure}

We derived the column densities using the {\it Owens} software obtained from the 
French FUSE team. The 
line minimum corresponds to the optically derived radial velocity. The typical 
virial velocity of 
BCGs of this luminosity is $<$ 100 kms$^{-1}$. The width of the absorption 
lines are much 
larger than this. This is probably due to outflows from the starburst and to a 
lesser extent 
also to infall of neutral gas. We have to adopt the fit of the line profiles to 
this condition and 
the programme we use conveniently allows one to fit lines belonging to several 
velocity 
components simultaneously. Our approach was to test a number of situations in 
which we assumed 
the cloud ensemble to consist of between 3 and 9 different velocity systems, in 
all cases 
covering the whole width of the lines. In the cases where the number of 
components were low this 
was compensated by the software by increasing the Doppler broadening parameter, 
{\it b}. The {\it b} parameter was allowed 
to vary between 30 and 180 kms$^{-1}$. FUSE has a resolution of $\sim$20 
kms$^{-1}$ and one might worry that there could be a significant number of 
unresolved saturated components with low {\it b} parameters in the sightline. 
From FUSE curve-of-growth of neutral and low ionization absorption line systems 
in the Magellanic bridge one obtains {\it b} parameters in the range 15-20 
kms$^{-1}$ (Sembach et al. \cite{sembach}, Lehner \cite{lehner}). The H$_2$ 
lines give smaller {\it b} values, indicating a stronger confinement and lower 
turbulence of these clouds. We do not expect the conditions in Haro 11 to be 
more quiet than in the Magellanic bridge and therefore we should be safe with a 
lower value of the {\it b} parameter of the low ionization lines of 30 
kms$^{-1}$. In the example we discuss below, however, we allow the parameter to 
vary freely, utilizing the full resolution of FUSE. In one setup of parameters 
we fixed the 
relative velocity 
differences between the different components to the same value,  so that the 
full width of 
the lines  was covered. We also tested fits where both the velocity differences 
and the line 
widths were allowed to vary freely. In all cases we assumed a gas temperature of 
300K. The absorption lines from all atoms and ions under study were included 
simultaneously in the fits. The optimum 
fits were found to contain between 5-7 components. Ideally one would expect the 
relative strengths of the components to be the same for each element. As the 
components get weaker and noiser however, the results become more and more 
uncertain. Thus while the central 3-4 components show a good agreement, the 
fainter components may deviate significantly from one species to the other. The 
goodness of fit (lowest 
total $\chi^2$ 
residual/numbers of degrees of freedom) showed small variations between fits to 
different numbers of subcomponents. As an example, 
one of best 
results for the 
carbon line is shown in Fig. \ref{3788fig8}.

\begin{table}[h]
\caption[]{Column densities of low ionization species}
\begin{flushleft}
\begin{tabular}{llll} 
Species & $\lambda$ & log N & $\sigma _{logN}$ \cr
\hline
O & 1039.2 & 15.48 & 0.02 \cr 
C & 1036.3 & 15.19 & 0.02 \cr 
Si & 1020.7 & 15.10 & 0.03 \cr 
Ar & 1048.2 & 13.67 & 0.03  \cr 
Fe & 1096.9 & 14.80 & 0.08 \cr
& 1125.4 & 14.59 & 0.13 \cr
\hline
\end{tabular}\\
\label{columndensities}
\end{flushleft}
\end{table}

From Table \ref{columndensities} we obtain an oxygen to carbon ratio of 
N(C)/N(O)$\sim$0.5. This agrees with the value derived from the IUE spectra (see 
sect. 2.1). Thus we feel confident that the procedure we applied above gives 
reliable results. From the fits we derived the column densities listed in Table 
\ref{columndensities}. We have 
previously obtained the oxygen abundance of H~II regions in Haro 11, 
log(N(O)/N(H))+12=7.9. We 
will assume that this is an upper limit of the oxygen abundance in the neutral 
region. Applying 
the conversion to hydrogen column densities from this number we obtain a lower 
limit to the 
hydrogen column density, logN(H~I)=19.6. We have no corresponding determination 
of the remaining 
elements but if we assume the relative abundances to be the same as the solar 
(Asplund et al., 
\cite{asplund2}) we obtain a lower limit to logN(H~I) of between 19.6 and 20.3 
with the lower 
value for the strong carbon line. These relative high values are not consistent 
with a diffuse 
distribution of H~I since it would prohibit leakage. We conclude that the 
photons must be leaking 
out through transparent windows in the gaseous halo in a "picket-fence fashion". 
This would also mean that 
the estimated 
column densities are slightly too low, due to the dilution. A possible scenario 
explaining why we 
observe a leakage is that the merging process has allowed young star clusters to 
separate from 
high density gaseous regions on a time scale shorter than the lifetime of the 
burst, 
i.e. in a short 
moment of strong gravitational disturbances.

In the following analysis we will assume that the picket-fence model is the best 
approximation of 
the situation.

\subsection{Extinction  corrections}

In the next section we will derive an estimate of the LyC 
escape fraction based on both the observed H$\alpha$ flux and the continuum 
fluxes at 900 and 960 \AA~rest wavelength 
of Haro 11, f$_{900}$ and f$_{960}$. The fluxes were integrated over spectral 
regions of widths 5 and 2 \AA~ respectively. As mentioned above, the 900~\AA~ 
continuum wavelength was chosen 
to allow a direct comparison with previous studies of Lyman leakage. The 
960~\AA~ wavelength was chosen for spectral evolutionary model comparisons since 
it represents a spectral region reasonably close to the Lyman limit but 
unaffected by Lyman absorption line wings and other strong lines.
The observed data need to be corrected for galactic and internal dust 
extinction. The 
galactic extinction 
in the direction of Haro 11 is A$_B$=0.049 (Schlegel et al. \cite{schlegel}). 
The 900 and 960~\AA~rest 
wavelengths are redshifted to $\sim$920 and 980 \AA~ in the observed spectrum. 
At these 
wavelength 
A$_B$=0.049 corresponds to A$_{920}$=0.177 and A$_{980}$=0.175 (Mathis 
\cite{mathis}). We will now discuss the internal extinction corrections.

In the picket-fence approximation, as is further detailed in the next section, 
we will assume that the LyC photons are escaping through dust free windows. The 
only goal for the remaining part of this section is thus to obtain a value of 
the extinction correction to the 960~\AA~region, A$_{960}$. From the 
H$\alpha$/H$\beta$ ratio of the central 4$\arcsec$x4$\arcsec$, corrected for 
underlying absorption 
(Bergvall and \"Ostlin \cite{bergvall4}), we obtain, assuming the Calzetti et 
al. 
(\cite{calzetti}) extinction law, an internal extinction in {\it B} of 
A$_B$=1.08$^m\pm$0.07$^m$. The 
FUSE aperture is 30$\arcsec \times$30$\arcsec$, i.e much larger than the 
aperture of the optical 
spectroscopy. The 
flux inside the 4$\arcsec \times$4$\arcsec$ aperture is only about 10\% of the 
total flux so an 
estimate of the 
global extinction based on these data is quite uncertain. If we assume that the 
dust column 
density is proportional to the {\it B} band flux we use the small aperture data 
to 
derive total 
extinction in {\it B} to be A$_B$=0.66$^m\pm$0.08$^m$. The error does not 
include 
the uncertainty 
of the model of the dust distribution. 

We can also estimate the extinction from the observed slope of the UV continuum 
in Haro 11, 
using a method developed by Calzetti et al. (\cite{calzetti}, \cite{calzetti2}) 
and Meurer et al. 
(\cite{meurer}). Calzetti et al. (\cite{calzetti}) investigated 39 starburst and 
blue compact 
galaxies to study the interstellar extinction in the UV. The observed spectral 
distribution in 
this region may be approximated by a power law F($\lambda$)$\propto \lambda 
^{\beta}$. They found 
that the optical depth obtained from the H$\alpha$/H$\beta$ ratio correlated 
with 
the slope 
$\beta$ and derived a value of $\beta$ in a dust free environment,  
$\beta_0$=-1.7. They also 
found that the optical depth calculated from the H$\alpha$/H$\beta$ ratio was a 
factor of 2 
larger than that derived from the continuum underlying the emission lines. This 
was interpreted 
as a consequence of a systematically higher extinction in young star forming 
regions as compared 
to older regions. 

Later, Meurer et al. (\cite{meurer}) found a correlation between $\beta$ and the 
ratio between 
the far infrared flux and the UV flux at 2200 \AA. This correlation agreed very 
nicely with a 
foreground dust screen model following the Calzetti et al. extinction law, if 
the galaxies had a 
similar intrinsic UV continuum distribution. The observed slope can therefore be 
used to derive 
an estimate of the extinction by comparing to the "standard" dust free slope. 
The value derived 
by Meurer et al. $\beta_0$=-2.5$\pm$0.5, is bluer than what Calzetti et al. 
found. This 
is probably due to the fact that the Meurer et al. sample contained more 
low--luminosity starburst 
dwarfs that were completely dominated by the young population while the more 
massive galaxies in 
the Calzetti et al. sample had a slightly larger contribution from older stars. 
The extinction of Haro 11 therefore depends on which value we choose. 
Haro 11 is involved 
in a very intense global starburst and is metal poor. On the other hand it is 
quite luminous and 
most likely has quite a complex star formation history so we expect a mixture of 
stellar 
populations of different ages. This puts Haro 11 somewhere in between the Meurer 
et al. sample and 
the Calzetti et al. sample. We will therefore assume that its dust free UV 
continuum is 
represented by the weighted mean of the Meurer et al. value and the Calzetti et 
al. value. Both 
have a similar observed scatter around the mean so the value we will adopt is 
$\beta_0$=-2.1$\pm$0.2. 

The UV slope in Haro 11, obtained from our IUE observations, covering the 
wavelength interval 
1250~\AA--1980~\AA~is $\beta$=-1.4$\pm$0.1. (In the picket-fence model it is 
incorrect to use the slope without correcting for the contribution from the 
unobscured part, but this would make a negligible difference here.) As an 
additional check of this value 
we can use the 
observed flux ratio F$_{FIR}$/F$_{UV}$ of Haro 11 and apply the model most 
favoured 
by Meurer et al. 
to calculate the predicted value of $\beta$. F$_{FIR}$ is derived from IRAS data 
and the 
approximation from Lonsdale \& Helou (\cite{lonsdale}). This results in
F$_{FIR}$=1.26$\cdot$10$^{-14}$(2.58F$_{60}$+F$_{100}$) W m$^{-2}$ = 
2.74$\cdot$10$^{-13}$ W m$^{-2}$. The flux at 2200 \AA~was obtained from an 
extrapolation of the 
fit to the shorter wavelength region. This gives us F$_{FIR}$/F$_{UV}$=5.9. From 
this value we 
obtain $\beta$=-1.6$\pm$0.1, agreeing nicely with the former value we derived. 
$\beta$=-1.4 
corresponds to an extinction in B of A$_B$=0.83, which is quite close to the 
value we obtain from 
the \hahb ratio (A$_B$=0.66). We will assume a value of A$_B$=0.7$\pm$0.2. 

The final value of the internal extinction was obtained in two steps. First we 
adopted the Calzetti et al. extinction law, relevant in the wavelength 
region 
1250 \AA--8000 \AA~, to determine the reddening correction at 1250 \AA. Then we 
used the recent determination of the extinction curve in the SMC by Cartledge et 
al. (\cite{cartledge}) to 
go from there to 960 \AA, the region we will use to sample the stellar 
continuum. We thus finally adopt 
an internal extinction  A$_{960}$=2.9$^m\pm$0.3. 

We will now, with the use of spectral evolutionary models, derive an estimate of 
the upper and lower 
flux escape limits.

\subsection{Estimate of the escape fraction}

\subsubsection{Comparison with spectral evolutionary models}
Under the assumption of a simple geometry for the distribution of stars, gas and
dust, limits on the global LyC escape fraction from Haro 11 may be inferred from
the observed LyC flux. In principle, the LyC leakage could be caused either by
holes in the gas (e.g. supernova chimneys), by a total gas mass too low to form
a complete Str\"omgren sphere, or by a combination thereof. Since the high
column densities of neutral hydrogen derived in Sect. 3.1 disfavour a truncated
Str\"omgren sphere for Haro 11, we will here estimate the escape fraction
assuming that the escape is exclusively due to an isotropic distribution of
holes, which are assumed to be devoid of gas (neutral and ionized) as well as
dust.

Following Deharveng et al. (\cite{Deharveng et al.}), we define the total LyC
escape fraction $f_\mathrm{esc}$ as
\begin{equation}
f_\mathrm{esc}=\frac{L_\mathrm{900,obs}}{L_{900,\star}},
\label{ezeq1}
\end{equation}
where $L_\mathrm{900,obs}$ represents the observed (escaping) luminosity 
(W \AA$^{-1}$) at a wavelength corresponding to 900 \AA~in the rest system of
the galaxy, and $L_{900,\star}$ represents the corresponding luminosity produced
by the stellar component before attenuation by gas. Under the approximations
adopted here, $L_{900,\star}$ is given by the sum of two components:
\begin{equation}
L_{900,\star}=L_\mathrm{900,obs}+L_\mathrm{900,abs}, 
\label{ezeq2}
\end{equation}
where $L_\mathrm{900,abs}$ represents the luminosity lost through gas absorption
inside the galaxy. $L_\mathrm{900,abs}$ is in turn related to the number of LyC
photons absorbed by gas $N_\mathrm{LyC,abs}$ through
\begin{equation}
L_\mathrm{900,abs}=N_\mathrm{LyC,abs}\times 10^{-k_1},
\label{ezeq4}
\end{equation}
where $k_1$ is a model-dependent parameter. Under the assumption of case B
combination, the number of absorbed LyC photons can then be related to the
extinction-corrected H$\alpha$ luminosity $L_{\mathrm{H}\alpha}$ through
\begin{equation}
N_\mathrm{LyC,abs}=\frac{\lambda_{\mathrm{H}\beta}}{hc(j_{\mathrm{H}\alpha}/j_{\
mathrm{H}\beta})}\frac{\alpha_\mathrm{B}}{\alpha^\mathrm{eff}_\mathrm{H\beta}}L_
{\mathrm{H}\alpha},
\label{ezeq5}
\end{equation}
where $j_{\mathrm{H}\alpha}/j_{\mathrm{H}\beta}$ is the intrinsic
$\mathrm{H}\alpha/\mathrm{H}\beta$ line ratio, $\alpha_\mathrm{B}$ is the Case B
recombination coefficient and $\alpha^\mathrm{eff}_\mathrm{H\beta}$ is the
effective recombination coefficient for the $\mathrm{H\beta}$ line. These three
parameters depend on the temperature of the nebula and, to lesser extent, on the
density. By assuming an electron density of 100 cm$^{-3}$ and fitting a cubic
spline to the $j_{\mathrm{H}\alpha}/j_{\mathrm{H}\beta}$, $\alpha_\mathrm{B}$
and $\alpha^\mathrm{eff}_\mathrm{H\beta}$ values given by Osterbrock
(\cite{Osterbrock}), we estimate that for the electron temperature of Haro 11
($T\approx 13400$ K; Bergvall \& \"Ostlin \cite{bergvall4}):
\begin{equation}
N_\mathrm{LyC,abs}\approx8.56\times 10^{11}L_{\mathrm{H}\alpha}.
\label{ezeq6}
\end{equation}
Combination of the equations above, with the relation $L=4\pi 
D^2 f$ between luminosity $L$, H$\alpha$ flux $f_{H\alpha}$, flux density 
$f_{\lambda}$ and distance $D$, gives:
\begin{equation}
f_\mathrm{esc}=\frac{f_{\mathrm{900,obs}}}{f_{\mathrm{900,obs}}+8.56\times 
10^{11-k_1}f_\mathrm{H\alpha}}.
\label{ezeq7}
\end{equation}
Constraints on $f_\mathrm{esc}$ may be derived from (\ref{ezeq7}) by combining
the estimated 1$\sigma$ error bars on $f_{900,\mathrm{obs}}$ and
$f_\mathrm{H\alpha}$ with extremum values of $k_1$ derived from models. When
evaluating the plausible range of the $k_1$ parameter, defined as 
\begin{equation}
k_1=\log (\frac{N_\mathrm{LyC,abs}}{L_\mathrm{900,abs}}),
\end{equation}
we use the Zackrisson et al. (\cite{Zackrisson et al.}, hereafter Z01) spectral 
evolutionary model. With this model, a grid of burst-like spectral evolutionary 
sequences is generated, defined by the model parameter values specified in in 
Table \ref{modelgrid}. In the case of a standard Salpeter IMF ($dN/dM\propto 
M^{-\alpha}$; $\alpha=2.35$ with upper mass limit $M_\mathrm{up}=120 \ M_\odot$) 
the range of $k_1$ values allowed by this grid of star formation scenarios 
becomes $k_1=13.23$--13.24. When the full range of IMF variations considered 
($\alpha=1.35$--3.35 and  $M_\mathrm{up}=20$--$120 \ M_\odot$) is used, this 
constraint relaxes to $k_1=12.70$--13.58.

When these values of $k_1$ are combined with the previously discussed
measurements of $f_\mathrm{900,obs}=1.06\pm 0.09 \times 10^{-17}$ W m$^{-2}$
\AA$^{-1}$ and $f_\mathrm{H\alpha}=4.5\pm 0.5\times10^{-15}$ W m$^{-2}$, limits 
on $f_\mathrm{esc}$ may be inferred, resulting in $0.04 \leq
f_\mathrm{esc} \leq 0.10$ for the standard IMF and $0.01 \leq
f_\mathrm{esc} \leq 0.11$ for the more generous range of IMF scenarios. 

These $f_\mathrm{esc}$ estimates rely heavily on the conversion from
$L_\mathrm{H\alpha}$ to $L_\mathrm{900,abs}$, and are only valid under the
assumption that the H$\alpha$ line is dominated by photoionization by stars,
i.e. that ionization contribution from shocks or an active galactic nucleus are 
negligible. We have also assumed the dust to be exclusively located outside the 
Str\"omgren sphere, so that no LyC photons are lost due to internal dust
extinction. Potential far-UV opacity sources outside Haro 11, other than dust in 
the Milky Way, have furthermore been neglected.

A consistency test may be carried out by comparing the observed slope of the 
spectrum across the Lyman limit, to the corresponding slope predicted by models. 
To do this, we measure the continuum flux density at a rest wavelength of 960 
\AA{}, where the spectrum should be completely unaffected
by gaseous absorption. After correction for Galactic extinction, the observed
flux density at this wavelength is $f_\mathrm{960,obs}=7.4\pm 0.4\times 
10^{-17}$ W
m$^{-2}$ \AA$^{-1}$. In sect. 3.2 the correction factor for internal extinction 
was estimated to be
$y_{960}=14^{+5}_{-3}$. Given the assumption that the holes in the nebula are
completely devoid of both gas and dust, and hence that the dust correction
applies only to the parts of the nebula where there is no LyC leakage, the
escape fraction may then be estimated from:
\begin{equation}
f_\mathrm{esc}= 
\frac{k_2f_\mathrm{900,obs}}{y_{960}f_\mathrm{960,obs}-(y_{960}-1)
k_2 f_\mathrm{900,obs}},
\label{ezeq9}
\end{equation}
where $k_2$ represents the ratio of the stellar component fluxes at 960 \AA{} to
900 \AA{} prior to absorption by gas and dust:
\begin{equation}
k_2=\frac{f_{960,\star}}{f_{900,\star}}.
\label{ezeq10}
\end{equation}
Although the exact value of $k_2$ is not known for Haro 11, a lower limit on 
$f_\mathrm{esc}$ can be imposed from the minimum $k_2$ predicted by the 
plausible range of star formation scenarios. From the model grid defined in 
Table \ref{modelgrid}, we infer $\min
(k_2)=1.78$ for the standard IMF and $\min
(k_2)=1.42$ otherwise. This translates into $f_\mathrm{esc}\geq
0.025^{+0.012}_{-0.010}$ and $f_\mathrm{esc}\geq
0.019^{+0.009}_{-0.003}$, respectively, which is completely consistent with the 
LyC escape fraction derived from the $\mathrm{H\alpha}$ flux. 

To further test the robustness of these results, two other spectral evolutionary 
models -- Starburst99 v. 4 (Leitherer et al. \cite{leitherer3}) and P\'EGASE.2 
(Fioc \& Rocca-Volmerange \cite{Fioc & Rocca-Volmerange}) -- have been used to 
assess the model-sensitivity of the $k_1$ and $k_2$ parameters. We find, that 
when identical input parameters are used in all codes, very similar results are 
produced, indicating that our limits on $f_\mathrm{esc}$ do not critically 
depend on the use of any specific model.

\begin{table}[h]
\caption[]{The grid of Z01 evolutionary sequences. The grid consists of all 
possible combinations of the parameter values listed below. Each evolutionary 
sequence runs from ages of 0 to 100 Myr in short time steps.}
\begin{flushleft}
\begin{tabular}{ll} 
\hline
IMF, $\alpha$ & 1.35, 2.35, 3.35 \cr
$M_{\mathrm{up}}$  & 20, 70, 120 $M_\odot$\cr
SFH & e10, c100 \cr
$Z_{\star} $ & 0.001 \cr
\hline
\end{tabular}\\
IMF: $dN/dM \propto M^{-\alpha}$\\
$M_\mathrm{up}$: The upper mass limit of the IMF\\
SFH=Star formation history.\\
\hspace{0.5cm}c=Constant star formation rate during the subsequent\\
\hspace{0.5cm} number of Myr.\\
\hspace{0.5cm}e=Exponentially declining star formation rate,\\
\hspace{0.5cm} SFR$(t)\propto\exp(-t/\tau)$, with an e-folding decay rate
($\tau$)\\ 
\hspace{0.5cm} equal to the subsequent number of Myr. \\
$Z_\star$: The metallicity of the stellar population\\
\label{modelgrid}
\end{flushleft}
\end{table}

\section{Discussion}

As we mentioned in the introduction it has been argued that star-forming 
galaxies may yield the 
required LyC emissivities to reionize the universe at $z>$5-6. This approach has 
the advantage 
over assuming AGNs as sources in that it explains the early pollution of heavy 
elements observed 
in the IGM at these redshifts (e.g. Cowie et al. \cite{cowie}). Moreover, 
analysis of 
the optical 
depth at the H~I and HeII \lya continuum breaks in different environments at 
$z<$4 indicates that 
the major ionization source in dense regions is QSOs while in void regions the 
radiation appears 
to be softer, giving room for a substantial contribution from starburst 
galaxies. 

Several semi-analytical methods have been used to model the reionization epoch. 
Due to the strong 
inhomogeneities of the radiation field in the early stages of the reionization, 
the computations 
are however quite demanding and complex (see Razoumov et al. \cite{razoumov} for 
a summary of the 
technical problems) and realistic simulations have not been possible until 
recently. In 2003,
Sokasian et al. (\cite{sokasian}) used a fast code capable of exploring the 
impact of low-mass 
starburst galaxies in high spatial resolution. They focused on the evolution of 
the interaction 
between the IGM and starforming galaxies of normal properties, e.g. similar to 
Haro 11, in the 
redshift interval $z\sim$20 to $z\sim$6. One of the results was that the 
previous estimates of 
the required photon flux rate needed to complete the reionization was 
overestimated by 63\% if 
assumed that only galaxies with masses $>$ 10$^9$h$^{-1}$\sm would account for 
the reionization. 
By including sources with masses down to $\sim$4$\times$10$^7$ h$^{-1}$\sm the 
impact of the large 
population of dwarf starburst galaxies could be realistically assessed. The 
calculations 
demonstrated that the soft ionization sources of type starburst dwarfs give the 
required photon 
flux provided that the escape fraction is $f_\mathrm{esc}\ge$0.20. This number 
is twice the upper limit of $f_\mathrm{esc,gas}$ that we derived for Haro 11. 
The estimate is however based on a normal stellar population. At redshifts 
z$>$10 pop III stars could dominate the population (e.g. Choudhury \& Ferrara 
\cite{choudhury}). In these stars 
the emissivity in the UV might be 
larger than in pop II by a factor 
of 2 (Tumlinson \& Shull \cite{tumlinson}; Bromm et al. \cite{bromm}). Thus, 
although $f_\mathrm{esc}$ would not change, the number of produced photons per 
unit mass would increase significantly. Thus it seems feasible that, at least at 
z$\gtrsim$10, dwarf 
starbursts of the 
mass of Haro 11 and 
lower could provide a substantial contribution to the early phase of the 
reionization process.

\section{Conclusions}

We report on the discovery of a weak signal in the Lyman continuum of the 
luminous starburst 
galaxy Haro 11. It is the first time Lyman continuum leakage is found in a 
galaxy in the local 
environment. Using the derived column densities of low ionization species we 
estimate the column 
density of neutral hydrogen to be $\sim$ 10$^{19}$ cm$^{-2}$, which is too high 
to allow Lyman 
continuum photon escape if the gas is diffusely distributed. We conclude that 
the 
Lyman continuum 
radiation is escaping through transparent windows of the ISM. Assuming a 
Salpeter IMF, we estimate a 
total escape fraction 
of 0.04$<f_\mathrm{esc}<$0.10. More extreme IMFs allow a lower escape fraction 
but the major uncertainty is due to the problematic background subtraction at 
short wavelengths. The upper limit is below the model estimates of the minimum 
escape rate required for reionization. Still, this single case demonstrates that 
a significant fraction of the required Lyman continuum photons may be produced 
by starburst dwarfs, in particular if we also consider that Pop III stars, that 
should dominate the first generation galaxies, are more efficient than Pop II 
stars in producing Lyman continuum photons.

\begin{acknowledgements}

This work has been done using the profile fitting procedure Owens.f developed by 
M. Lemoine and the FUSE French Team. We are indebted to Carmelle Robert for help 
with the access to the LavalSB synthetic spectral library. We are pleased to 
thank our referee, Daniel Kunth, for many valuable comments and suggestions to 
improvements of the manuscript. Kjell Olofsson is thanked for his 
observational contribution to the IUE data. This work was supported by the Swedish Science COuncil and the Swedish Space Board.

\end{acknowledgements}

\end{document}